\documentclass[12pt]{IOS-Book-Article-CPA-2011}
\usepackage{times}
\usepackage{mathptmx}
\usepackage[cmex10]{amsmath}
\usepackage{
    amssymb, 
    amsthm, 
    amsfonts,
    latexsym,
    array,
    graphicx,
    comment,
    url,
    alltt,
}
\usepackage[caption=false,font=footnotesize]{subfig}
\usepackage[bookmarks, bookmarkstype=toc]{hyperref}

\allowdisplaybreaks

\makeatletter
\newbox\sf@box
\newenvironment{SubFloat}[2][]
  {\def\sf@one{#1}
   \def\sf@two{#2}
   \setbox\sf@box\hbox
   \bgroup}
 { \egroup
  \ifx\@empty\sf@two\@empty\relax
    \def\sf@two{\@empty}
  \fi
  \ifx\@empty\sf@one\@empty\relax
    \subfloat[\sf@two]{\box\sf@box}%
  \else
    \subfloat[\sf@one][\sf@two]{\box\sf@box}%
  \fi}
\makeatother

\newenvironment{packeditemize}
{\begin{itemize}
    \setlength{\itemsep}{1pt}
    \setlength{\parskip}{0pt}
    \setlength{\parsep}{0pt}}
{\end{itemize}}
\newenvironment{packedenumerate}
{\begin{enumerate}
    \setlength{\itemsep}{1pt}
    \setlength{\parskip}{0pt}
    \setlength{\parsep}{0pt}}
{\end{enumerate}}

\newcommand{\ben}{\begin{enumerate}}
\newcommand{\een}{\end{enumerate}}
\newcommand{\bit}{\begin{itemize}}
\newcommand{\eit}{\end{itemize}}
\newcommand{\bpen}{\begin{packedenumerate}}
\newcommand{\epen}{\end{packedenumerate}}
\newcommand{\bpit}{\begin{packeditemize}}
\newcommand{\epit}{\end{packeditemize}}
\newcommand{\bvb}{\begin{verbatim}}
\newcommand{\evb}{\end{verbatim}}

\newcommand{\exnotn}[1]{$$\text{#1}$$}

\newenvironment{egnotn}
{\begin{minipage}[t]{\textwidth}\vspace{0ex}\begin{quote}\begin{alltt}\normalfont}
{\end{alltt}\end{quote}\vspace{0ex}\end{minipage}}

\newcommand{\fig}[1]{Figure~\ref{fig:#1}}

\newcommand{\sect}[1]{Section~\ref{sec:#1}}

\newcommand{\thmref}[1]{Theorem~\ref{thm:#1}}

\newcommand{\eqnref}[1]{Equation~\ref{eqn:#1}}

\newcommand{\kw}[1]{{\bf #1}}

\newcommand{\brkt}[1]{\left(#1\right)}
\newtheorem{theorem}{Theorem}
\newtheorem{corollary}[theorem]{Corollary}

\newcommand{\thrm}[1]{\begin{theorem}#1\end{theorem}}
\newcommand{\prf}[1]{\begin{proof}#1\end{proof}}

\theoremstyle{definition}
\newtheorem{definition}{Definition}
\newcommand{\defn}[1]{\begin{definition}#1\end{definition}}

\newcommand{\procdistribute}{{\small \textsf{distribute}~}}
\newcommand{\procsmsort}{{\small \textsf{seq-msort}~}}
\newcommand{\procpmsort}{{\small \textsf{par-msort}~}}
\newcommand{\procnode}{{\small \textsf{node}~}}
\newcommand{\procmerge}{{\small \textsf{merge}~}}

\DeclareMathOperator*{\len}{length}
\DeclareMathOperator*{\jump}{jump}
\DeclareMathOperator*{\tmin}{T_s^{m}}

\normalfont
\begin{document}

\begin{frontmatter}                           
\setcounter{page}{1}

\title{Fast Distributed Process Creation with the XMOS XS1 Architecture}
\runningtitle{Fast Distributed Process Creation with the XMOS XS1 Architecture}

\author{\fnms{James} \snm{HANLON}}~~and~~\author{\fnms{Simon} \snm{J.} \snm{HOLLIS}}
\address{Department of Computer Science, University of Bristol, UK.\\%
{\small \{{\tt hanlon}~,~{\tt hollis}\}~{\tt @cs.bris.ac.uk}}}
\runningauthor{J.\ Hanlon and S.\ J.\ Hollis}

\begin{abstract} 
The provision of mechanisms for processor allocation in current
distributed parallel programming models is very limited.  This makes difficult,
or even prohibits, the expression of a large class of programs which require a
run-time assessment of their required resources.  This includes programs whose
structure is \emph{irregular}, \emph{composite} or \emph{unbounded}.
Efficient allocation of processors requires a \emph{process creation} mechanism
able to initiate and terminate remote computations quickly.
This paper presents the design, demonstration and analysis of an explicit
mechanism to do this, implemented on the XMOS XS1 architecture, as a foundation
for a more dynamic scheme.  It shows that process creation can be made efficient
so that it incurs only a fractional overhead of the total runtime and that it
can be combined naturally with recursion to enable rapid distribution of
computations over a system.  
\end{abstract}

\end{frontmatter}

\section*{Introduction}

An essential issue in the design of scalable, distributed parallel computers is
the rate at which computations can be initiated, and results collected as they
terminate \cite{May99}. This requires an efficient method of \emph{process
creation} capable of dispatching a program and data on which to operate to a
remote processor. This paper presents the design, implementation, demonstration
and evaluation of a process creation mechanism for the XMOS XS1 architecture
\cite{XS1}.

Parallelism is being employed on an increasingly large scale to improve
performance of computer systems, particularly in high performance systems, but
increasingly in other areas such as embedded computing \cite{Asanovic06}. As
current programming models such as MPI (Message Passing Interface) provide
limited support for automated management of processing resources, the burden of
doing this mainly falls on the programmer. These issues are not relevant to the
expression of a program as, in general, a programmer is concerned only with
introducing parallelism (execution on multiple processors) to improve
performance, and not how the computation is scheduled on the underlying system.
When we consider that future high performance systems will run on the order of
$10^9$ threads \cite{Exascale10}, it is clear that the programming model must
provide some means of dynamic processor allocation to remove this burden. This
is the situation we have with memory in sequential systems, where allocation and
deallocation is performed with varying degrees of automaticy.

This observation is not new \cite{May88, Hansen90}, but it is only as existing
programming models and software struggle to meet the increasing scale of
parallelism that the problem is again coming to light. For instance,
capabilities for \emph{process creation and management} were introduced in the
MPI-2.0 specification, stating that: ``\emph{Reasons for including process
management in MPI are both technical and practical.  Important classes of
message-passing applications require this control. These include task farms,
serial applications with parallel modules and problems that require a run-time
assessment of the number and type of processes that should be started}''
\cite{MPI2Spec}.
Several MPI implementations support process creation and management
functionality, but it is pitched as an `advanced' feature that is difficult to
use and problematic with many current job-scheduling systems. More
encouragingly, language-level abstractions for dynamic process creation and
placement have appeared recently in the Chapel \cite{Chapel} and X10 \cite{X10},
which are being developed by Cray and IBM respectively as part of DARPA's High
Productivity Computing Systems program. Both support these concepts as key
ingredients in the design of parallel programs, but they are built on software
communication libraries and statically-mapped program binaries.  Consequently,
they are subject to the same communication inefficiencies and inflexibility of
single-program approaches. 

A run-time assessment of required processing resources concerns large class of
programs whose structure is \emph{irregular}, such as unstructured-grid
algorithms like the Spectral Element Method \cite{Patera84}, \emph{unbounded}
such as recursively-structured algorithms like Branch-and-Bound search
\cite{BranchAndBound} and Adaptive Mesh Refinement \cite{Berger1984}, or
\emph{composite}, where a program may be composed of different parallel
subroutines that are themselves executed in parallel, possibly each with its own
structure.
These all require a means of dynamic processor allocation that is able to
distribute computations \emph{over} a set of processors, depending on
requirements determined at runtime. The combination of parallelism and recursion
is a powerful mechanism for \emph{growth} which can be used to implement
distribution efficiently. This must be supported with a mechanism for process
creation with the ability to dispatch, initiate and terminate computations
efficiently on remote processors.

This paper presents the design and implementation of an explicit scheme for
dynamic process creation in a distributed memory parallel computer.  This work
is intended to be a key building block for a more automatic scheme.
The implementation is on the the XMOS XS1 architecture, which has low-level
provisions for concurrency, allowing a convincing proof-of-concept
implementation. 
Based on this, the process creation mechanism is evaluated by combining it with
controlled recursion in two simple algorithms to demonstrate the \emph{rate} and
\emph{granularity} at which it is possible to create remote computations. Performance
models are developed in each case to interpret the measured results and to make
predictions for larger systems and workloads. 
This analysis highlights the efficiency, scalability and effectiveness of the
concept and approach taken.



The rest of this paper is structured as follows. \sect{background} describes the
XS1 architecture, the experimental platform and the notations and conventions
used. \sect{implementation} gives a brief overview of the design and
implementation details. \sect{evaluation} presents the performance models and
experimental and predicted results. Finally, \sect{conclusion} concludes and
\sect{future} discusses possible future extensions to the work.

\section{Background}\label{sec:background}

\subsection{Platform}

The XMOS XS1 processor architecture \cite{XS1} is general-purpose,
multi-threaded, scalable and has been designed from the ground up to support
concurrency. It allows systems to be constructed from multiple \emph{XCore}
processors which communicate with each other through fast communication links.
The key novel aspect of this architecture with respect to the work in this paper
is the instruction set support for processes and communication. Low-level
threading and communication are key features, exposed with operations, for
example, to provide synchronous and asynchronous fork-join thread-level
parallelism and channel-based message passing communication.  Provision of these
features in hardware allows them to be performed in the same order of magnitude of
time as memory references, branches and arithmetic. This allows efficient
high-level notations for concurrency to be effectively built.

The system used to demonstrate and evaluate the proposed process creation
mechanism is an experimental board called the XK-XMP-64 \cite{XMP64}. It
connects together 64 XCore processors in 16 XS1-G4 devices which run at 400MHz.
The G4 devices are interconnected in a 4-dimensional hypercube which
equivalently can be viewed as a 2-dimensional torus. Mathematically, this is
defined in the following way \cite{Leighton91}:

\begin{definition} A $d$-dimensional hypercube is a graph $G=(N,E)$ where $N$ is
the set of $2^d$ nodes and $E$ is the set of edges. Each node is labeled with a
$d$-bit identifier. For any $m, n\in N$, an edge exists between $m$ and $n$ if
and only if $$m \oplus n = 2^k$$ for $0 \leq k \leq d$ where $\oplus$ is the
bitwise exclusive-or operator. Hence, each node has $d=\log N$ edges and
$|E|=d2^{d-1}$.\end{definition}

Each core in the G4 package has a private 64kB memory and is interconnected via
internal links to an integrated switch. It is convenient to view the whole
system as a 6-dimensional hypercube. As each core can run 8 hardware threads,
the system is capable of 512-way concurrency with an aggregate 25.6 GIPS
performance. 

\subsection{Notation}

For presentation of the algorithms in this paper, a simple imperative,
block-structured notation is used.  The following points describe the
non-standard elements that appear in the examples.

\subsubsection{Sequential and Parallel Composition} 

A set of instructions that are to be executed in sequence are composed with the
`;' separator. A sequence of instructions comprises a \emph{process}. For
example, the block \exnotn{\{ $I_1$ ; $I_2$ ; $I_3$ \}} defines a simple process
to perform three instructions, $I_1$, $I_2$ and $I_3$ in sequence. Processes may
be executed in parallel by composition within a block with the `$|$' separator.
Execution of a parallel block initiates the execution of the constituent
processes simultaneously. The parallel block successfully terminates only when
all processes have successfully terminated. This is referred to as synchronous
fork-join parallelism.  For example, the block declaration \exnotn{\{ $P_1$ $|$
$P_2$ $|$ $P_3$ \}} denotes the parallel execution of three processes $P_1$,
$P_2$ and $P_3$.

\subsubsection{Aliasing} 

The \textbf{aliases} statement is used to create new
references to sub-sections of an array. For example, the statement \exnotn{A
\kw{aliases} B[\(i\dots j\)]} sets A to refer to the sub-section of B in the index
range $i$ to $j$.

\subsubsection{Process Creation} 

The \textbf{on} statement reveals explicitly to the programmer the process
creation mechanism. The statement \exnotn{\kw{on} $p$ \kw{do} $P$} is
semantically equivalent to executing a call to $P$, except that process $P$ is
transmitted to processor $p$, which then executes $P$ and communicates back any
results using channels, leaving the original processor free to perform other
tasks. By composing \textbf{on} in parallel, we can exploit multi-threaded
parallelism to offload work while executing another process. For example, the
statement \exnotn{\{ $P_1$ $|$ \kw{on} $p$ \kw{do} $P_2$ \}} causes $P_1$ to be
executed while $P_2$ is offloaded and executed on processor $p$.

\subsection{Measurements}

All timing measurements presented were made with hardware timers, which are
accessible through the ISA and have 10ns resolution. Constant values were
extrapolated through the measurements taken by fitting performance models to the
data.

\subsection{Conventions}

All logarithms are to the base 2. $p$ is defined as the number of processors and
is taken to be a positive power of two. A word is taken to be 4 bytes and is
a unit of input in the performance models.

\section{Implementation}\label{sec:implementation}

The \textbf{on} statement causes the \emph{closure} of a process $P$ located at a
\emph{guest} processor to be sent to a remote \emph{host} processor, the host to
execute $P$ and to send back any updated free variables of $P$ stored at the
guest.  The execution of \textbf{on} is synchronous in this respect. The closure
of a process $P$ is a complete description of $P$ allowing it to be executed
independently and is defined in the following way:

\defn{The closure $C$ of a process $P$ consists of three elements: a set of
arguments $A$, which represents the complete variable context of $P$ as we don't
consider global variables, a set of procedure indicies $I$ and a set of
procedures $Q$: $$C(P) = (A, I, Q)$$ where $|A| \geq 0$ and $|I| = |Q| \geq 1$.
Each argument $a \in A$ is a ordered sequence of one or more integer values.
Each process $P \in Q$ is an ordered sequence of one or more instructions. $I_P$
is an integer value denoting the index of procedure $P$.} 

Each core maintains a fixed-size \emph{jump table} denoted `$\jump$', which
records the location of each procedure in memory. As the procedure address may
not be consistent between cores the indicies are guaranteed to be. This allows
relative branches to be expressed in terms of an index which is locally
referenced at execution.  Each node in the system is initialised with a minimal
binary containing the process creation kernel. The complete program is loaded on
node 0, from where parts of it can be copied onto other nodes to be executed.

\subsection{Protocol}

The process creation mechanism is implemented as a point-to-point protocol
between a \emph{guest} core and a \emph{host} core.  Any running thread is able
to spawn the execution of a process on any other core.  It consists of the
following four phases.

\subsubsection{Connection Initialisation}

A guest initiates a connection by sending a single byte control token and a word
identifying itself. It waits for an acknowledgment from the host indicating a
host thread has been allocated and the connection is properly established. A
core may host multiple guest computations, each on a different thread.

\subsubsection{Transmission of Closure}

$C(P)$ is transmitted in three parts.  Firstly, a header is sent containing
$|A|$ and $|Q|$.  Secondly, each $a \in A$ is sent with a single word header
denoting the type of the argument. For referenced arrays, this is followed by
$\len(a)$ and the values contained. The host writes these directly into
heap-allocated space and the argument value is set to this address. Single-value
variables are treated similarly and constant values can be copied directly into
the argument value.  Lastly, each $P \in Q$ is sent with a two word header
denoting $I_P$ and $\len(P)$ in bytes. The host allocates space on the heap and
receives the instructions of $P$ from the guest, read from memory in word-chunks
from $\jump[I_P]$ to $\jump[I_P] + \len(P)$. On completion, the host sets
$\jump[I_P]$ to the address of $P$ on the heap.

\subsubsection{Execution/Wait for Completion}

Once $C$ has been successfully transmitted, the host initialises the thread's
registers and stack with the arguments of $P$ and initiates execution.
The connection is left open and the guest thread waits for the host to indicate
$P$ has halted.

\subsubsection{Transmission of Results and Teardown}

Once $P$ has halted, all referenced array and variable arguments contained in
$C$ (now the results) are transmitted back to the guest. The guest writes them
back directly to their original locations.  Once this has been completed, the
connection is terminated. The guest continues execution and the host thread
frees the memory allocated to the closure and yields.

\subsection{Performance Model}

The runtime cost of this mechanism is captured in the following way:
\defn{The runtime of process creation $T_c$ is a function of the total size of
the argument values $n$, procedure descriptions $m$ and the results $o$ and is
given by $$T_c(n,m,o) = (C_i + C_w n + C_w m + C_w o) \cdot C_l$$ where $C_i$
and $C_w$ are constants relating to initialisation and termination, and overhead
per (word) value transmitted respectively. The value $n$ is inclusive of the
size of referenced arrays and hence $o \leq n$. As all communication is
synchronised, $C_l$ is a constant factor overhead relating to the latency of the
path between the guest and host processors.}
Normalising $C_l=1$ to a single hop off-chip, the per-word overhead $C_w$ was
measured as 150ns. The initialisation overhead $C_i$ is dependent
on the size of the closure.

\section{Demonstration and Evaluation}\label{sec:evaluation}

The aim of this section is to \emph{demonstrate} the use of process creation
combined with parallel recursion to \emph{evaluate} the performance of the
design and its implementation in realising efficient growth. To do this, we
develop performance models to combine with experimental results, allowing us to
extrapolate to larger systems and inputs. 
We start with a simple algorithm to demonstrate the fast distribution of
parallel computations and then show how this can be applied to a practical
problem.

\subsection{Rapid Process Distribution}


The algorithm \procdistribute given in \fig{dist} is inspired by \cite{May99}
and works by spawning a new copy of itself on a remote processor each time it
recurses. Each process then itself recurses, continuing this behaviour and
hence, each level of the recursion subdivides the set of processors in half,
resulting in a doubling of the capacity to initiate computations. This growth
follows the structure of a binary tree. When each instance of \procdistribute
executes with $n = 1$, the \emph{node} process is executed and the recursion
halted. The parameter $t$ indicates the node identifier and the algorithm is
executed from node 0 with $t=0$ and $n=p$.

\begin{figure}[t!]
\begin{egnotn}
\kw{proc} \procdistribute(\(t\), \(n\)) \kw{is}
    \kw{if} \(n=1\) \kw{then} \procnode(\(t\))
    \kw{else} 
    \{ \procdistribute(\(t\), \(n/2\))
    \(|\) \kw{on} \(t+n/2\) \kw{do} \procdistribute(\(t+n/2\), \(n/2\)) \}
\end{egnotn}

\caption{A recursive process \procdistribute to rapidly distribute another
process \procnode over a set of processors.}

\label{fig:dist}
\end{figure}

\subsubsection{Runtime}

The hypercube interconnection topology of the XK-XMP-64 provides an optimal
transport in terms of hop distance between remote creations; this is established
by the following theorem.

\thrm{Every copy of \procdistribute is always created on a neighbouring node
when executed on a hypercube. \label{thm:hop}}

\prf{Let $H=(N,E)$ be a $d$-dimensional hypercube. When \procdistribute is
executed with $t=0$ and $n=N$, starting at node 0 on $H$, the recursion follows
the structure of a binary tree of depth $d=\log |N|$, where identifiers at level
$i$ are multiples of $|N|/2^i$. A node $p$ at depth $i$ with identifier
$k|N|/2^i$ creates a new \emph{remote} child node $c$ with identifier
$k|N|/2^i+|N|/2^{i+1}$. As $|N|=2^d$, $c = k2^{d-i}+2^{d-i-1}$ and hence, $p
\oplus c = 2^{d-i-1}$.}

Given that $m$ and $n$ are fixed, that $o=0$ (there are no results) and from
\thmref{hop} we can normalise $C_l$ to 1, the runtime $T_{c}(m,n,o)$ of the
\textbf{on} statement in \procdistribute is $\Theta(1)$, which we define as the
initialisation overhead $C_j$. Using this, we can
express the parallel runtime of \procdistribute $T_{d}$ on $p$ processors. 
In each step, the
number of active processes double, but we count the runtime at each level of
recursion, which terminates when $n/2^i=1$ or $i=\log n$. Hence, 
\begin{align}
T_d(p) = & \sum_{i=1}^{\log p} \left ( T_c + C_o
\right)\nonumber\\
= & (C_j + C_o)\log p \label{eqn:tdist}
\end{align}
where $C_o$ is the the sequential overhead at each level. $C_j$ was measured as
18.4$\mu$s and $C_o$ was measured as 60ns.



\subsubsection{Results}

\fig{distResults} gives the predicted and measured execution time of
\procdistribute as a function of the number of processors. The prediction
\emph{almost exactly} matches the runtime given by \eqnref{tdist}.
\fig{distLevelResults} shows the inaccuracy between the measured and predicted
results more clearly, by giving the measured execution time for each level in
the recursion, that is, the difference between consecutive points in
\fig{distResults}. It shows that the assumption made based on \thmref{hop} does
not hold and that the first two levels take fractionally less time than the last
four levels (3.85$\mu$s). This is due to the reduced on-chip communication
costs. Overall though, each level of recursion completes on average in
18.9$\mu$s and it takes only 114.60$\mu$s to populate all 64 processors. Moreover,
using the performance model given by $T_d$, we can extrapolate to larger $p$
than is possible to measure with the current platform.  For example, when
$p=1024$, $T_d(1024)$ = 190$\mu$s.

\begin{figure}
  \begin{minipage}[b]{0.49\linewidth}
    \centering
    \subfloat[Measured vs. predicted ($\star$) execution time.]{
      {\small\input{figures/distribute_.tex}}
      \label{fig:distResults}
    }
  \end{minipage}
  \begin{minipage}[b]{0.49\linewidth}
    \centering
    \subfloat[Execution times for each level of recursion of \procdistribute.]{
      {\small\input{figures/distributeLevels_.tex}}
      \label{fig:distLevelResults}
    }
\end{minipage}

\caption{Measured execution time of \procdistribute over varying numbers of
processors. \protect\subref{fig:distLevelResults} clearly shows the inter- vs.
intra-chip latencies.}

\end{figure}

\subsubsection{Remarks}

By using the performance model to make predictions, we have assumed a hypercube
topology and efficient support for concurrency. Although other architectures and
larger systems cannot make such provisions, the model and results provide a
reasonable lower bound on execution time with respect to the approach described. 

The hypercube has rich communication properties and supports exponential growth,
but it does not scale well due to the number of connections at each node and
length of wires in realistic packagings. Although \procdistribute has optimal
single-hop behaviour and we obtain peak performance, it is well known that
efficient embeddings of binary trees into lower-degree networks such as meshes
and tori exist \cite{Leighton91}, allowing reasonable dispersion. In this case,
the granularity of process creation would have to be chosen to match the
capabilities of the architecture. 

Provision of efficient ISA-level operations for processes and communications
allows fine-grained performance, particularly in terms of short messages. Many
current architectures do not support these operations at a such a low-level and
cannot exploit the full potential of this approach, although again it
generalises at a coarser granularity of message size to match the relative
performance of these operations. 

\subsection{Mergesort}

Mergesort is a well known sorting algorithm \cite{Mergesort} that works by
recursively halving a list of unsorted numbers until unit sub-lists are
obtained. These are then successively \emph{merged} together such that each
merging step produces a sorted sub-list, which can be performed in time
$\Theta(n)$ for sub-lists of size $n/2$. \fig{seqmergesort} gives the sequential
mergesort algorithm \procsmsort.

Mergesort's branching recursive structure matches that of \procdistribute,
allowing us to combine them to obtain a parallel version.  Instead of
sequentially evaluating the recursive calls, conditional on some threshold value
$C_{th}$, a local recursive call is made in parallel with the second call which
is migrated to a remote core. This threshold is used to control the extent to
which the computation is distributed.  In each of the experiments for an input
of size $2^k$ and available processors $p=2^d$, the threshold is set as $2^k/p$.
The approach taken in \procdistribute is used to control the placements of
each of the sub-computations.  Initially, the problem is split in half; this
will have the greatest benefit to the execution time. Depending on the problem
size, further remote branchings of the problem may not be economical, and the
remaining steps should be evaluated locally, in sequence. In this case, the
algorithm simply reduces to \procsmsort.

This parallel formulation of mergesort is essentially just \procdistribute
with additional work and communication overhead, but it will allow us to more
concretely quantify the relative costs of process creation. The parallel
implementation of mergesort \procpmsort is given in \fig{parmergesort}.  It
uses the same sequential \procmerge procedure and the parameters $t$ and $n$
control the placement of processes in the same way as they were used with
\procdistribute.

\begin{figure}
\begin{SubFloat}[]{\label{fig:seqmergesort} 
}
\begin{minipage}[c]{0.47\linewidth}
\begin{egnotn}
\kw{proc} \procsmsort(\(A\)) \kw{is}
  \kw{if} \(|A| > 1\) \kw{then}
  \{ \(a \kw{aliases} A[0..|A|/2-1]\)
  ; \(b \kw{aliases} A[i..|A|]\)
  ; \procsmsort(\(a\))
  ; \procsmsort(\(b\))
  ; \emph{merge}(\(A\),\(a\),\(b\))
  \}
\end{egnotn}
\end{minipage}
\end{SubFloat}
\begin{SubFloat}[]{\label{fig:parmergesort} 
}
\begin{minipage}[c]{0.47\linewidth}
\begin{egnotn}
\kw{proc} \procpmsort(\(t\), \(n\), \(A\)) \kw{is}
  \kw{if} \(|A| > 1\) \kw{then}
  \{ \(a \kw{aliases} A[0\dots|A|/2-1]\)
  ; \(b \kw{aliases} A[i\dots|A|]\)
  ; \kw{if} \(|A| > C\sb{th}\) \kw{then}
    \{ \procpmsort(\(t\), \(n/2\), \(a\))
    \(|\) \kw{on} \(t+n/2\) \kw{do}
            \procpmsort(\(t+n/2\), \(n/2\), \(b\)) \}
    \kw{else}
    \{ \procpmsort(\(t\), \(n/2\),\(a\))
    ; \procpmsort(\(t+n/2\), \(n/2\),\(b\)) \}
  ; \emph{merge}(\(A\),\(a\),\(b\))
  \}
\end{egnotn}
\end{minipage}
\end{SubFloat}
\caption{Sequential and parallel mergesort processes.}
\end{figure}

We can now analyse the performance and behaviour of \procpmsort and the
process creation mechanism by looking at the parallel runtime.

\subsubsection{Runtime}

We first define the runtime of the sequential components of \procpmsort.
This includes the sequential merging and sorting procedures. The runtime $T_m$ of
\procmerge is linear and is defined as $$T_m(n) = C_a n + C_b$$ for constants $C_a, C_b >
0$, relating to the per-word and per-\procmerge overheads respectively. 
These were measured as $C_a=90$ns and $C_b=830$ns.  The runtime $T_s(n,1)$
of \procsmsort, is expressed as a recurrence:
\begin{equation}
T_s(n,1) = 2T_s\brkt{\frac{n}{2},1} + T_{m}(n)\label{eqn:tseqmsortrecc}
\end{equation}
which has the solution
\begin{equation}
T_s(n,1) = n(C_c\log n+C_d)\label{eqn:tseqmsort}
\end{equation}
for constants $C_c, C_d > 0$. These were measured as $C_c=200ns$ and
$C_d=1200ns$. Based on this we can express the runtime of \procpmsort as
the combination of the costs of creating new processes, moving data, merging and
sorting sequentially. The key component of this is the cost $T_{c}$, relating to
the \textbf{on} statement in the parallel formulation, which is defined as
$$T_{c}(n)=C_i + 2C_w n.$$ This
is because we can normalise $C_l$ to 1 (due to \thmref{hop}), the size of the
procedures sent is constant and the number of arguments and results are both
$n$. The initialisation overhead $C_i$ was measured as 28$\mu$s, larger than
that for \procdistribute as the closure contains the descriptions of \procmerge
and \procpmsort. For the parallel runtime, the base sequential case is given by
\eqnref{tseqmsortrecc}.  With two processors, the work and execution time can be
split in half at the cost of migrating the procedures and data:
$$T_s(n,2) = T_{c}\brkt{\frac{n}{2}} + T_s\brkt{\frac{n}{2},1} + T_{m}(n).$$ With
four processors, the work is split in half at a cost of $T_{c}(n/2)$ and then in
quarters at a cost of $T_{c}(n/4)$. After the data has been sequentially sorted
in time $T_s(n/4,1)$ it must be merged at the two children of the master node in
time $T_{m}(n/2)$, and then again at the master in time $T_{m}(n)$: 
\begin{align*}
T_s(n,4) = & T_{c}\brkt{\frac{n}{2}} + T_{c}\brkt{\frac{n}{4}}
+ T_{m}\brkt{\frac{n}{2}} + T_{m}(n) + T_s\brkt{\frac{n}{4},1}
\end{align*}
Hence in general, we have:
\begin{align}
T_s(n,p) = & \sum_{i=1}^{\log p} \brkt{
T_{c}\brkt{\frac{n}{2^i}} + T_{m}\brkt{\frac{n}{2^{i-1}}} } 
+ T_s \left ( \frac{n}{p},1 \right )\nonumber
\end{align}
for $n \geq p$ as each leaf sub-process of the sorting computation must operate
on at least one data item. We can then express this precisely by substituting
our definitions for $T_s$, $T_c$ and $T_m$ and simplifying:
\begin{align}
T_s(n,p) = & C_w\frac{2n}{p}(p-1) + C_i \log p + C_a\frac{2n}{p}(p-1) 
+ C_b \log p + \frac{n}{p}\brkt{C_c \log \frac{n}{p} + C_d}\nonumber\\
= & \frac{2n}{p}(p-1)(C_w+C_a) + (C_i + C_b)\log p
+ \frac{n}{p}\brkt{C_c \log \frac{n}{p} + C_d} \label{eqn:tsimple}
\end{align}
For $p=1$, this reduces to \eqnref{tseqmsort}. This definition allows us to
express the a lower bound and minimum for the runtime.

\subsubsection{Lower Bound}

We can give a lower bound $\tmin$ on the parallel runtime $T_s(n,p)$ such
that $\forall n,p$ $$T_s(n,p) \geq \tmin .$$ This is obtained by considering
the \emph{parallel overhead}, that is the cost of distributing the problem over
the system. In this case it relates to the cost of process creation, including
moving processes and their data, the $T_c$ component of $T_s$:
\begin{align}
\tmin(n,p) & = \sum_{k=1}^{\log p} T_{c}\brkt{\frac{n}{2^k}}\nonumber\\
& = \sum_{k=1}^{\log p} \left ( C_i + 2C_w \frac{n}{2^k} \right )\nonumber\\
& = C_i \log p + C_w\frac{2 n}{p} (p-1) \label{eqn:tlowerbound}.
\end{align}
\eqnref{tlowerbound} is then the sum of the costs of process creation and
movement of input data. When $n=0$, $\tmin$ relates to \eqnref{tdist}; this is 
the cost of transmitting and initiating just the computations over the system.
For $n \geq 0$, this includes the cost of moving the data.

\subsubsection{Minimum}

Given an input of length $m\leq n$ for some sub-computation of \procpmsort,
creation of a remote branch is beneficial only when the cost of this is less
than the local sequential case:
\begin{align*}
T_{c}\brkt{\frac{m}{2}} + T_s\brkt{\frac{m}{2},1} + T_{m}(n) & < 
    T_s(m,1)\\
T_{c}\brkt{\frac{m}{2}} + T_s\brkt{\frac{m}{2},1} + T_{m}(n) & <
    2T_s\brkt{\frac{m}{2},1} + T_{m}(m)\\
T_{c}\brkt{\frac{m}{2}} & < T_s\brkt{\frac{m}{2},1}
\end{align*}
Hence, initiation of a remote sorting process for an array of length $n$ is
beneficial only when $$T_{c}(n) < T_s(n,1).$$ That is, the cost of
remotely initiating a process to perform half the work and receiving the results
is less than the cost of sequentially sorting $m/2$ elements.  Therefore at the
inflection point we have 
\begin{equation}
T_{c}\brkt{n} = T_s\brkt{n,1}\label{eqn:optimalpayoff}.
\end{equation}


\subsubsection{Results}

\begin{figure*}
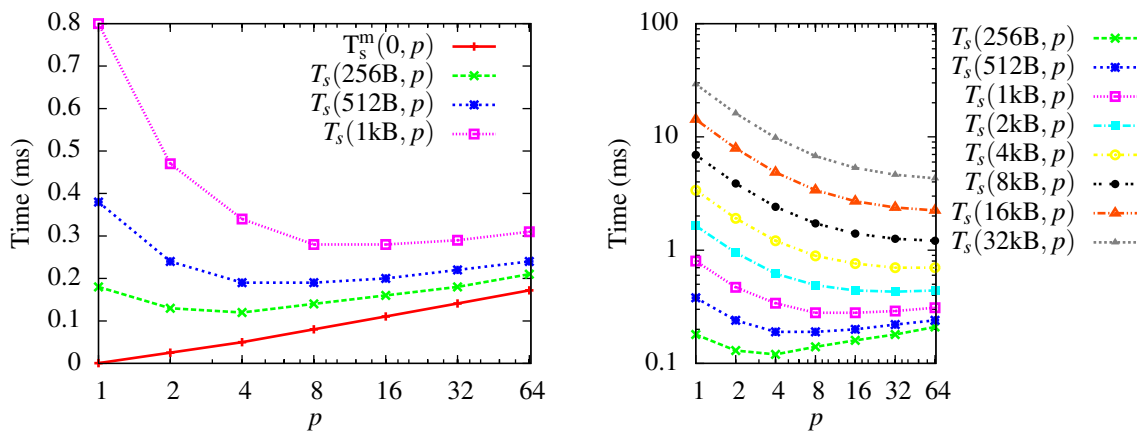

\begin{minipage}[b]{0.49\linewidth}
  \centering
  \subfloat[Log-linear plot for varying small inputs.]{
    {\small\input{figures/time2_.tex}}
    \label{fig:time1}
  }
\end{minipage}
\hfill
\begin{minipage}[b]{0.49\linewidth}
  \centering
  \subfloat[Log-log plot for larger inputs.]{
    {\small\input{figures/time1_.tex}}
    \label{fig:time2}
  }
\end{minipage}

\caption{Measured execution time of \procpmsort as a function of the
number of processors. \protect\subref{fig:time1} highlights the minimum
execution time and the $\tmin$ lower bound.}

\label{fig:measuredmsort}
\end{figure*}

\fig{measuredmsort} shows the measured execution time of \procpmsort as a
function of the number of processors used for varying input sizes. \fig{time1}
shows just three small inputs.  The smallest possible input is 256 bytes as the
minimum size for any sub-computation is 1 word. The minimum execution time for
this size is at $p=4$ processors, when the array is subdivided twice into 64
byte sections.  This is the point given by \eqnref{optimalpayoff} and indicates
directly the total cost incurred in offloading a computation. For $p<4$, the
cost of sorting sequentially dominates the runtime, and for $p>4$, the cost of
creating a new processes and transferring the array sections dominates the
runtime.  With the next input of size 512 bytes, the minimum moves to $p=8$,
where the array is again divided into 64 byte sections. This holds for each
input size and in general gives us the minimum size for which creating a new
process will further reduce the runtime. 

The runtime lower bound $\tmin (0,p)$ given by \eqnref{tlowerbound} is also
plotted on \fig{time1}. This shows the small and sub-linear cost with
respect to $p$ of the overheads incurred with the distribution and management
of processes around the system. Relative to $T_s(64,p)$ this constitutes most of
the overall work performed, which is expected as the array is fully decomposed
into unit sections. For larger sized inputs, as presented in \fig{time2}, this
cost becomes just a fraction of the total work performed.

\begin{figure*}
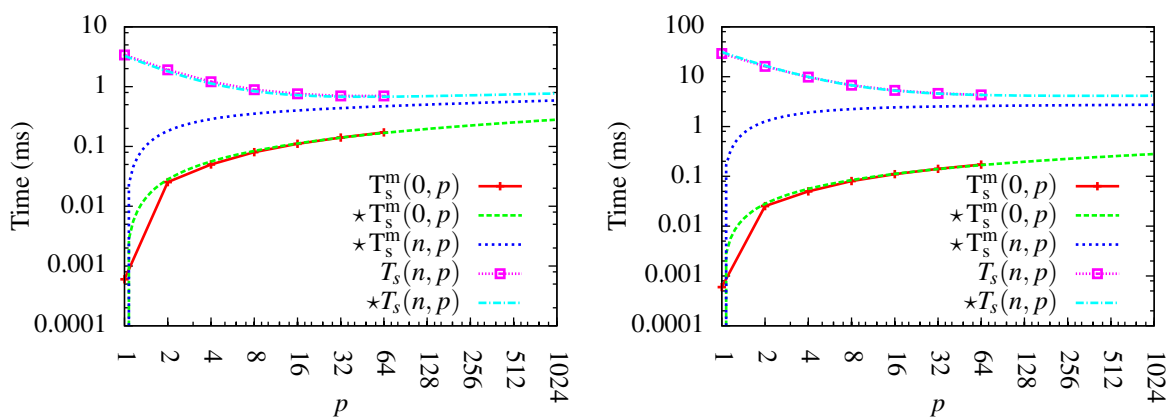
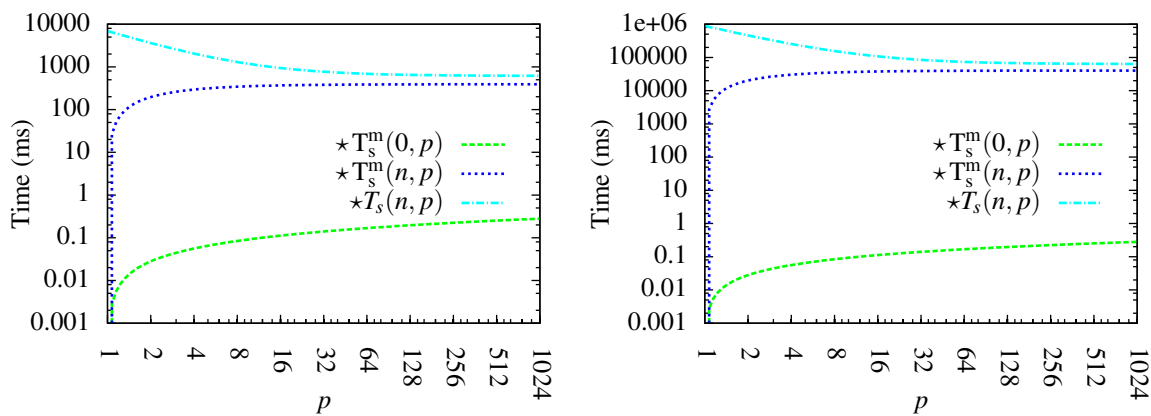

  \begin{minipage}[b]{0.49\linewidth}
    \centering
    \subfloat[$n=64$ (256B) with measured results up to 64 cores.]{
      {\small\input{figures/pred1_.tex}}
      \label{fig:pred1}
    }
  \end{minipage}
  \hfill
  \begin{minipage}[b]{0.49\linewidth}
    \centering
    \subfloat[$n=8192$ (32kB) with measured results up to 64 cores.]{
      {\small\input{figures/pred2_.tex}}
      \label{fig:pred2}
    }
  \end{minipage}
  \begin{minipage}[b]{0.49\linewidth}
    \centering
    \subfloat[$n=2621440$ (10MB).]{
      {\small\input{figures/pred3_.tex}}
      \label{fig:pred3}
    }
  \end{minipage}
  \hfill
  \begin{minipage}[b]{0.49\linewidth}
    \centering
    \subfloat[$n=268435465$ (1GB).]{
      {\small\input{figures/pred4_.tex}}
      \label{fig:pred4}
    }
  \end{minipage}

\caption{Predicted ($\star$) performance of \procpmsort for larger $n$
and $p \leq 1024$. All plots are log-log.}

\label{fig:predictedmsort}
\end{figure*}

\fig{predictedmsort} shows predicted execution times for \procpmsort for
larger $p$ and $n$. Each plot contains the execution time $T_s$ as defined by
\eqnref{tsimple}, and $\tmin$ with and without the transfer of data. \fig{pred1}
gives results for the smallest input size possible to sort on 1024 cores (4kB)
and includes the measurements for $\tmin(0,p)$ and $T_s$. It reiterates what was
shown in \fig{time1} and shows that beyond 64 cores, very little penalty is
incurred to create up to 1024 sorting instances, with $\tmin$ accounting for
around 23\% of the total runtime for larger systems. This is due to the
exponential growth of the distribution mechanism. \fig{pred2} gives results for
the largest measured input of 32kB, showing the same trends, where $\tmin$ this
time is around just 3\% of the runtime between 64 and 1024 cores.

\fig{pred3} and \fig{pred4} present predictions made by the performance model
for more \emph{realistic} workloads of 10MB and 1GB respectively. \fig{pred3}
shows that 10MB could be sorted sequentially in around 7s and in parallel in
at least 0.6s. \fig{pred4} shows that 1GB could be sorted in just under 15m
sequentially or at least 1m in parallel. 
What these results make clear is that the distribution of the input
data dominates and bounds the runtime and that the distribution of data
constituting the process descriptions is a negligible proportion of the overall
runtime for reasonable workloads.
The relatively small sequential workload $O(n/p \log (n/p))$ of mergesort, which
decays quickly as $p$ increases, emphasises the cost of data distribution. For
heavier workloads, such as $O((n/p)^2)$, we would expect to see a much more
dramatic reduction in execution time and the cost of data distribution still
eventually to bound runtime, but then by a relatively fractional amount.

\section{Conclusions}\label{sec:conclusion}

This paper presents the design, implementation, demonstration and
evaluation of an efficient mechanism for dynamically creating computations in a
distributed memory parallel computer. It has shown that a computation can be
dispatched to a remote processor in just tens of microseconds, and when this
mechanism is combined with recursion, it can be used to efficiently implement
parallel growth.

The \procdistribute algorithm demonstrates how an \emph{empty} array of
processors can be populated with a computation exponentially quickly. For 64
cores, it takes just 114.60$\mu$s and for 1024 cores this will be of the order
of 190$\mu$s. 
The \procpmsort algorithm extends this by performing additional
computational work and communication of data which allowed us to obtain a
clearer picture of the cost of process creation with respect to varying problem
sizes. As the cost of transferring and invoking remote computations is related
primarily to the size of the closure, this cost grows slowly with system size
and is independent of data. With a 10MB input, it represents around
just 0.001\% of the runtime. 

The sorting results also highlight two important issues: the granularity at
which it is \emph{possible} to create new processes and costs of data movement.
They show that the computation can be subdivided to operate on just 64 byte
chunks and for performance to still be improved.  The cost of data movement is
significant, relative to the small amount of work performed at each node; for
more intensive tasks, these costs would diminish. However, these results assume
a worst case, where all data originates from a single core. In other systems,
this cost may be reduced by concurrent access through a parallel file system or
from prior data distribution.

The XS1 architecture provides efficient support for concurrency and
communications and the XK-XMP-64 provides an optimal transport for the described
algorithms, so we expect our lightweight scheme to be \emph{fast}, relative to
the performance of other distributed systems.  Hence, the results provide a
convincing proof-of-concept implementation, demonstrating the kind of
performance that is possible and, with respect to the topology, establish a
reasonable lower bound on the performance of the approach presented.
The results generalise to more dynamic schemes where placements are not perfect
and other larger architectures such as supercomputers, where interconnection
topologies are less well connected and communication is less efficient. In these
cases, the approach applies at a coarser granularity with larger problem sizes
to match the relative performance.

\section{Future Work\label{sec:future}}

Having successfully designed and implemented a language and runtime allowing
explicit process creation with the \textbf{on} statement, we will continue with
our focus on the concept of growth in parallel programs and plan to extend the
work in the following ways.  
Firstly, by looking at how placement of process closures can be determined
automatically by the runtime, relieving the programmer of having to specify
this. 
Secondly, by implementing the language and runtime with C and MPI to target a larger
platform, which will provide a more scalable demonstration of the concepts and their
generality.
And lastly, by looking at generic optimisations that can be made to the
process creation mechanism to improve overall performance and scalability.
More details about the current implementation are available
online\footnote{\url{http://www.cs.bris.ac.uk/~hanlon/sire}}, where news of
future developments will also be published.

\section*{Acknowledgments}

The authors would like to thank XMOS for their support, in particular from David
May, Henk Muller and Richard Osborne.

\bibliographystyle{unsrt}
{\small\bibliography{refs}}

\begin{thebibliography}{10}

\bibitem{May99}
David May.
\newblock {The Transputer revisited}.
\newblock In {\em {Millennial Perspectives in Computer Science: Proceedings of
  the 1999 Oxford-Microsoft Symposium in Honour of Sir Tony Hoare}}, pages
  215--246. Palgrave Macmillan, 1999.

\bibitem{XS1}
David May.
\newblock {\em The XMOS XS1 Architecture}.
\newblock XMOS Ltd., October 2009.
\newblock \url{http://www.xmos.com/support/documentation}.

\bibitem{Asanovic06}
{Asanovic, Bodik et al.}
\newblock {The Landscape of Parallel Computing Research: A View from Berkeley}.
\newblock Technical Report UCB/EECS-2006-183, EECS Department, University of
  California, Berkeley, Dec 2006.
\newblock
  \url{http://www.eecs.berkeley.edu/Pubs/TechRpts/2006/EECS-2006-183.html}.

\bibitem{Exascale10}
{Dongarra, J., Beckman, P. et al.}
\newblock {International Exascale Software Project Roadmap}.
\newblock Technical Report UT-CS-10-654, University of Tennessee EECS Technical
  Report, May 2010.
\newblock \url{http://www.exascale.org/}.

\bibitem{May88}
D.~May.
\newblock {The Influence of VLSI Technology on Computer Architecture [and
  Discussion]}.
\newblock {\em Philosophical Transactions of the Royal Society of London.
  Series A, Mathematical and Physical Sciences}, 326(1591):pp. 377--393, 1988.

\bibitem{Hansen90}
Per~Brinch Hansen.
\newblock The nature of parallel programming.
\newblock {\em Natural and Artifical Parallel Computation}, pages 31--46, 1990.

\bibitem{MPI2Spec}
{MPI} 2.0.
\newblock Technical report, Message Passing Interface Forum, November 2003.
\newblock \url{http://www.mpi-forum.org/docs/}.

\bibitem{Chapel}
B.L. Chamberlain, D.~Callahan, and H.P. Zima.
\newblock Parallel programmability and the {C}hapel language.
\newblock {\em International Journal of High Performance Computing
  Applications}, 21(3):291--312, 2007.

\bibitem{X10}
Philippe Charles, Christian Grothoff, Vijay Saraswat, Christopher Donawa, Allan
  Kielstra, Kemal Ebcioglu, Christoph von Praun, and Vivek Sarkar.
\newblock {X10}: an object-oriented approach to non-uniform cluster computing.
\newblock In {\em OOPSLA '05: Proceedings of the 20th annual ACM SIGPLAN
  conference on Object-oriented programming, systems, languages, and
  applications}, pages 519--538, New York, NY, USA, 2005. ACM.

\bibitem{Patera84}
A.~Patera.
\newblock {A spectral element method for fluid dynamics: Laminar flow in a
  channel expansion}.
\newblock {\em Journal of Computational Physics}, 54(3):468--488, June 1984.

\bibitem{BranchAndBound}
Bernard Gendron and Teodor~Gabriel Crainic.
\newblock Parallel branch-and-bound algorithms: Survey and synthesis.
\newblock {\em Operations Research}, 42(6):1042--1066, 1994.

\bibitem{Berger1984}
Marsha~J Berger and Joseph Oliger.
\newblock Adaptive mesh refinement for hyperbolic partial differential
  equations.
\newblock {\em Journal of Computational Physics}, 53(3):484 -- 512, 1984.

\bibitem{XMP64}
XMOS.
\newblock {\em XK-XMP-64 Hardware Manual}.
\newblock XMOS Ltd., Feburary 2010.
\newblock \url{http://www.xmos.com/support/documentation}.

\bibitem{Leighton91}
F.~Thomson Leighton.
\newblock {\em Introduction to parallel algorithms and architectures: array,
  trees, hypercubes}.
\newblock Morgan Kaufmann Publishers Inc., San Francisco, CA, USA, 1992.

\bibitem{Mergesort}
D.~E. Knuth.
\newblock {\em The Art of Computer Programming}, volume 3, Sorting and
  Searching, chapter 5.2.4, Sorting by Merging, pages 158--168.
\newblock Reading, MA: Addison-Wesley, 2nd ed. edition, 1998.

\end{thebibliography}

\end{document}